\documentclass[twocolumn,aps]{revtex4-1}
\usepackage{lineno,amsmath,bm,color,array,graphicx,url,xcolor,hyperref}

\newcommand{\pa}{\partial}
\renewcommand{\=}{\!=\!}
\newcommand{\1}{^{\mbox{\tiny (1)}}}
\newcommand{\2}{^{\mbox{\tiny (2)}}}
\newcommand{\n}{^{\mbox{\tiny (n)}}}
\newcommand{\B}[1]{{\bm{#1}}}
\newcommand{\C}[1]{{\mathcal{#1}}}
\newcommand{\h}{_{\mbox{\tiny h}}}

\hypersetup{
   colorlinks,
   linkcolor={blue!100!black},
   citecolor={green!50!black},
   urlcolor={blue!80!black}
}

\begin{document}

\title{Critical nucleation length for accelerating frictional slip}
\author{Michael Aldam$^{1}$, Marc Weikamp$^{2}$, Robert Spatschek$^{2}$, Efim A. Brener$^{3}$, and Eran Bouchbinder$^{1}$}
\affiliation{$^{1}$ Chemical Physics Department, Weizmann Institute of Science, Rehovot 7610001, Israel\\
$^{2}$ Institute for Energy and Climate Research, Forschungszentrum J\"ulich, J\"ulich, Germany\\
$^{3}$ Peter Gr\"unberg Institut, Forschungszentrum J\"ulich, D-52425 J\"ulich, Germany}


%

\begin{abstract}
The spontaneous nucleation of accelerating slip along slowly driven frictional interfaces is central to a broad range of geophysical, physical and engineering systems, with particularly far-reaching implications for earthquake physics. A common approach to this problem associates nucleation with an instability of an expanding creep patch upon surpassing a critical length $L_c$. The critical nucleation length $L_c$ is conventionally obtained from a spring-block linear stability analysis extended to interfaces separating elastically-deformable bodies using model-dependent fracture mechanics estimates. We propose an alternative approach in which the critical nucleation length is obtained from a related linear stability analysis of homogeneous sliding along interfaces separating elastically-deformable bodies. For elastically identical half-spaces and rate-and-state friction, the two approaches are shown to yield $L_c$ that features the same scaling structure, but with substantially different numerical pre-factors, resulting in a significantly larger $L_c$ in our approach. The proposed approach is also shown to be naturally applicable to finite-size systems and bimaterial interfaces, for which various analytic results are derived. To quantitatively test the proposed approach, we performed inertial Finite-Element-Method calculations for a finite-size two-dimensional elastically-deformable body in rate-and-state frictional contact with a rigid body under sideway loading. We show that the theoretically predicted $L_c$ and its finite-size dependence are in reasonably good quantitative agreement with the full numerical solutions, lending support to the proposed approach. These results offer a theoretical framework for predicting rapid slip nucleation along frictional interfaces.
\end{abstract}
\maketitle

\section{Introduction}

The process of rupture nucleation in which slowly driven frictional interfaces (faults) spontaneously develop elastodynamically propagating fronts accompanied by rapid slip is of fundamental importance for various fields, with far-reaching implications for earthquake physics. Quantitatively understanding the nucleation process is essential for predicting the dynamics of frictional interfaces in general and for earthquake dynamics in particular. There exists some observational evidence, based on seismological records~\citep{Scholz1998,Ohnaka2000,Harris2017}, and some experimental evidence, based on laboratory measurements~\citep{Dieterich1979,Ohnaka1990,Kato1992,McLaskey2013,Latour2013}, which suggest that rapid rupture propagation accompanied by a marked seismological signature is preceded by precursory aseismic slip. This precursory aseismic slip is commonly associated with a slowly expanding creep patch defined as a slipping segment of finite linear size $L(t)$, embedded within a non-slipping fault. Accelerating slip is expected to emerge once $L(t)$ surpasses a critical nucleation length $L_c$. We note that other nucleation scenarios have been considered in the literature, see for example~\citet{Ben-Zion2008}, but are not discussed here.

Various theoretical and computational works have indicated that the nucleation of accelerating slip is related to a frictional instability~\citep{Ruina1983,Yamashita1991,Ben-Zion1997,Scholz1998,Ben-Zion2001,Lapusta2003,Uenishi2003,Ben-Zion2008,Kaneko2008,Kaneko2016}. From this perspective, the critical nucleation length $L_c$ corresponds to the critical conditions for the onset of instability that leads to accelerating slip and to the spontaneous propagation of elastodynamic rupture fronts. A major challenge is to understand the relations between the critical instability conditions and $L_c$. In this Letter, we propose a theoretical approach for predicting $L_c$ which differs from the conventional approach.

The conventional approach, based on a single degree-of-freedom spring-block analysis extended to deformable bodies using various model-dependent fracture mechanics estimates, is discussed in the framework of rate-and-state constitutive laws in Sect.~\ref{sec:conventional}. Our approach, based on the stability of homogeneous sliding of elastically-deformable bodies, is introduced in Sect.~\ref{sec:LSA} and is shown to yield a significantly larger $L_c$ for elastically identical half-spaces and rate-and-state friction. In Sect.~\ref{sec:bimaterial} we show that the proposed approach is naturally applicable to bimaterial interfaces, which are of great interest in various contexts~\citep{Weertman1980,Andrews1997,Ben-Zion1998,Cochard2000,Adams2000,Ben-Zion2001,Gerde2001,Ranjith2001,Rice2001,Shi2006,Rubin2007,Ampuero2008a,allam2014,Brener2016,Aldam2017}, and derive analytic results for $L_c$ in this case, indicating that the bimaterial effect decreases $L_c$ compared to available predictions in the literature. Finally, in Sect.~\ref{sec:FiniteH} we show that the proposed approach is applicable to finite-size systems and test our predictions against inertial Finite-Element-Method calculations for a finite-size two-dimensional elastically-deformable body in rate-and-state frictional contact with a rigid body under sideway loading. The theoretically predicted $L_c$ and its finite-size dependence are shown to be in reasonably good quantitative agreement with the full numerical solutions, lending support to the proposed approach. Section~\ref{sec:conclusion} offers some concluding remarks and discusses some prospects.

\section{A conventional approach to calculating the nucleation length $L_c$}
\label{sec:conventional}

As stated, the most prevalent approach to the nucleation of rapid slip at frictional interfaces associates nucleation with an instability of a slowly expanding creep patch. The creep patch features a non-uniform spatial distribution of slip velocity, in the quasi-static regime (where inertia and acoustic radiation are negligible), due to some external loading. It  is assumed to be stable as long as its length $L(t)$ is smaller than a critical nucleation length $L_c$. When $L(t)\=L_c$, the patch becomes unstable and transforms into a rupture front, accompanied by accelerated slip and dynamic propagation (where inertia and significant acoustic radiation are involved). As creep patches are non-stationary objects that involve spatially varying fields, determining their stability --- and hence $L_c$ --- is a non-trivial challenge that typically requires invoking some approximations.

The most common approximation proceeds in two steps~\citep{Dieterich1986,Dieterich1992,Lapusta2000,Kaneko2008}. First, the creep patch and the two elastically deformable bodies that form the frictional interface are replaced by a rigid block of mass $M$ in contact with a rigid substrate and attached to a Hookean spring of stiffness $K$. That is, all of the spatial aspects of the problem are first neglected. The external loading and the typical slip velocity within the patch are mimicked by constantly pulling the Hookean spring at a velocity $V$. The rigid block is pressed against the rigid substrate by a normal force $F_N$, which gives rise to a frictional resistance force $f F_N$, where $f$ is described by the friction law, which may depend on the block's slip $u(t)$, its time-derivatives and the state of the frictional interface.

This single degree-of-freedom spring-block system is described by the force balance equation $M \ddot{u}(t)\=K(V t-u(t))-f(...)F_N,$ where each superimposed dot denotes a time-derivative. We assume that $f(...)$ can be described by the rate-and-state constitutive framework, where $f(\dot{u}(t),\phi(t))$ is a function of the slip velocity $\dot{u}$ and of an internal state variable $\phi$. The latter, which quantifies the typical age/maturity of contact asperities, evolves according to $\dot{\phi}\=g(\phi\,\dot{u}/D)$, where $D$ is a memory lengthscale and the function $g(\Omega)$ satisfies $g(1)\=0$ and $g'(1)\!<\!0$. For example, two popular choices, i.e.~$g(\Omega)\=1-\Omega$~\citep{Ruina1983, Marone1998,Nakatani2001,Baumberger2006,Bhattacharya2014} and $g(\Omega)\=-\Omega\log\Omega$~\citep{Ruina1983,Gu1984,Bhattacharya2014}, feature $g'(1)\=-1$.

Consider then a steady sliding state at a constant driving velocity $\dot{u}\=V$ such that $\phi\=D/V$. A standard linear stability analysis implies that this steady state becomes unstable if~\citep{Rice1983,Ruina1983,Gu1984,Lapusta2000,Baumberger2006,Bhattacharya2014}
\begin{equation}
\label{eq:spring-block}
K < K_c\equiv\frac{df(V,D/V)}{d\!\log{V}}\frac{g'(1)\,F_N}{D}\ ,
\end{equation}
where an inertial term proportional to $MV^2$ has been neglected. That is, an instability is predicted when the spring stiffness $K$ is smaller than a critical stiffness $K_c$. Note that since generically $g'(1)\!<\!0$, a necessary condition for instability is $df(V,D/V)/dV\!<\!0$, i.e.~that the sliding velocity $V$ belongs to the velocity-weakening branch of the steady state friction curve~\citep{Ruina1983}.

In the second step, the analysis is extended to spatially varying fields and elastically deformable bodies --- relevant to realistic creep patches --- by identifying the spring stiffness $K$ in the spring-block system with an $L$-dependent effective stiffness $K^{eff}\!(L)$ in the spatially varying and elastically deformable system. This is typically done through some fracture mechanics estimates which yield~\citep{Dieterich1986,Rice1993}
\begin{equation}
K^{eff}\!(L)=\eta\frac{\mu A_n}{L}\ ,
\end{equation}
where $\mu$ is the shear modulus, $A_n$ is the nominal contact area and the dimensionless number $\eta$ is a model-dependent pre-factor. As expected physically, the effective stiffness of the overall system, $K^{eff}$, is a decreasing function of the length of the creep patch, $L$. Using then $K^{eff}\!<\!K_c$ of Eq.~\eqref{eq:spring-block} as an instability criterion, one obtains
\begin{equation}
\label{eq:1DOF_Lc}
L > L_c \equiv \eta\frac{\mu D}{\tfrac{df(V,D/V)}{d\!\log{V}}g'(1)\,\sigma_0}\ ,
\end{equation}
where $\sigma_0\!=\!F_N/A_n$. The numerical pre-factor $\eta$ is model-dependent (e.g.~it depends on the crack configuration, dimensionality and loading configuration) and its value varies between $2/\pi$ and $4/3$ in the available literature~\citep[see Table 1]{Dieterich1992}. The nucleation criterion in Eq.~\eqref{eq:1DOF_Lc}, with $\eta$ close to unity, is widely used in the literature, though we are not aware of computational or experimental studies that quantitatively and systematically tested it. Next, we present a different approach for calculating $L_c$.

\section{An approach based on the stability of homogeneous sliding of elastically-deformable bodies}
\label{sec:LSA}

Our goal here is to propose an alternative approach to calculating the critical nucleation length $L_c$. In the proposed approach, nucleation is viewed as a spatiotemporal instability occurring along the creep patch which is assumed to be stable from the fracture mechanics perspective, i.e.~to propagate under stable Griffith energy balance conditions~\citep{Freund1990}. Since, in general, an elastic body can be thought of as a scale-dependent spring, one expects short wavelength $\lambda$ (large wavenumber $k\=2\pi/\lambda$) perturbations to be stable and instability --- if it exists --- to emerge beyond a critical (minimal) wavelength $\lambda_c$ (i.e.~below a critical wavenumber $k_c$). Consequently, when the size $L(t)$ of the expanding creep patch is small, $L(t)\!<\!2\pi/k_c$, we expect it to be stable. A loss of stability is expected when an unstable perturbation can {\em first} fit into the creep patch, i.e.~when the patch size satisfies $L(t)\=L_c\!\equiv\!2\pi/k_c$.

In this physical picture, the major goal is to calculate the critical wavenumber $k_c$. There is, however, no unique and general procedure to study the stability of non-stationary (time-dependent) and spatially varying solutions such as those associated with an expanding creep patch. Consequently, we invoke an approximation in which the spatially varying slip velocity within the creep patch is replaced by a homogeneous (space-independent) characteristic slip velocity $V$. With this approximation in mind, we need to study the stability of steady-state homogeneous sliding of an infinitely long system (in the sliding direction) in order to calculate $k_c$. Applying the result to the actual creep patch, accelerating slip nucleation is predicted to occur when $L(t)\=L_c\!\equiv\!2\pi/k_c$. This idea has been introduced, pursued and substantiated in the context of thin layers sliding on top of rigid substrates in~\citet{Bar-Sinai2013}. Our aim here is to significantly generalize the idea to any frictional system.
\begin{figure*}[ht]
  \centering
  \includegraphics[width=0.8\textwidth]{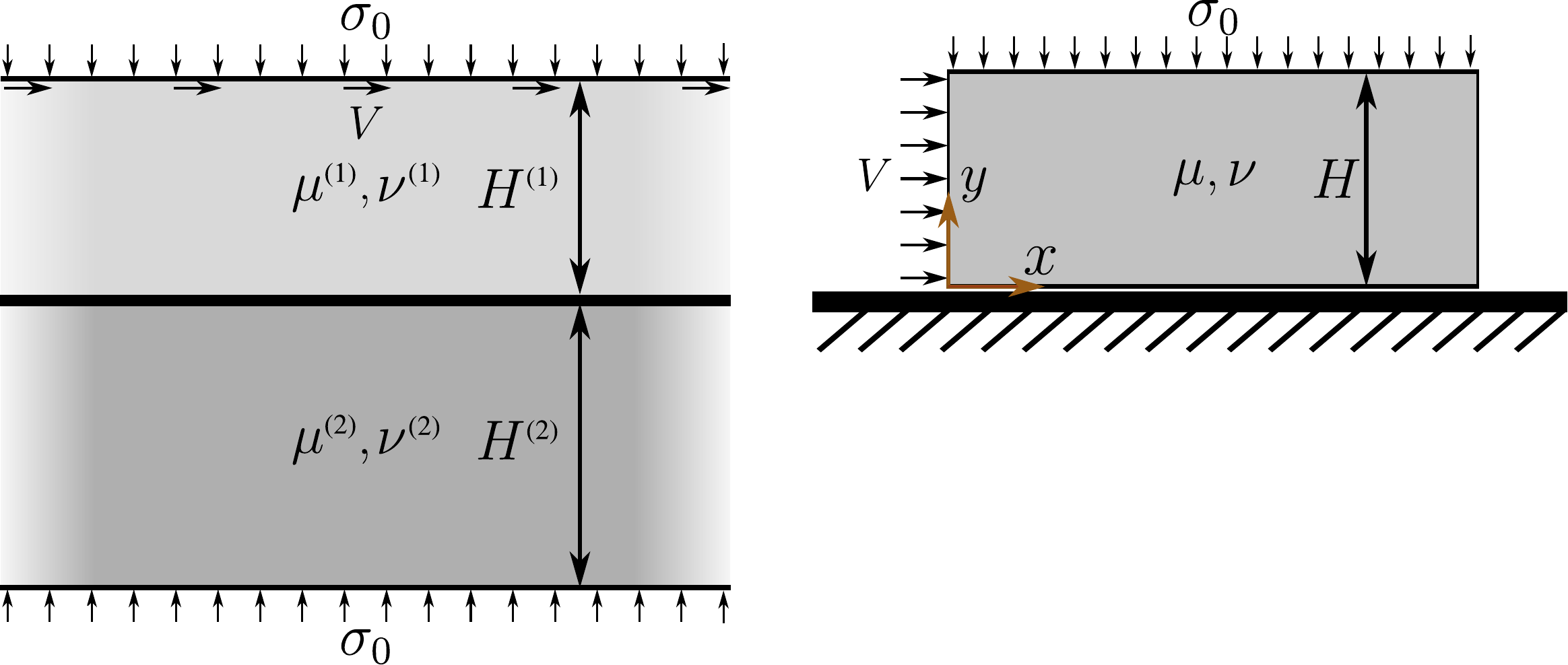}\\
  \caption{(left) A long elastic body of height $H\1$, shear modulus $\mu\1$ and Poisson's ratio $\nu\1$ sliding on top of another long elastic body of height $H\2$, shear modulus $\mu\2$ and Poisson's ratio $\nu\2$. The color gradients represent the fact that the bodies are essentially infinitely long. The bodies are pressed one against the other by a normal stress of magnitude $\sigma_0$ and a homogeneous sliding state at a relative velocity $V$ (in the figure the lower body is assumed to be stationary) is reached by the application of a shear stress of magnitude $\tau_0$ to the top and bottom edges (not shown). (right) The same as in the left panel, except that the lower body is infinitely rigid, $\mu\2\!\to\!\infty$, the upper body is of finite length and the velocity $V$ is applied to the lateral edge at $x\!=\!0$. Note that the superscript $\hbox{\scriptsize (1)}$ is unnecessary here and hence is omitted.}
  \label{fig:sys_fig}
\end{figure*}

We consider a long elastic body in the $x$-direction of height $H\1$ in the $y$-direction steadily sliding with a relative slip velocity $V$ on top of a long elastic body of height $H\2$. The bodies may be made of different elastic materials and are pressed one against the other by a normal stress $\sigma_0$, see Fig.~\ref{fig:sys_fig} (left). As we are interested in the response of the system to spatiotemporal perturbations on top of the homogeneous sliding state at a velocity $V$, we define the slip displacement $\epsilon(x,t)\!\equiv\! u_x(x,y\!=\!0^+,t)-u_x(x,y\!=\!0^-,t)$ and the slip velocity $v(x,t)\!\equiv\!\dot{\epsilon}(x,t)$, where ${\B u}(x,y,t)$ is the displacement field and $y\=0$ is the fault plane (the superscript $\hbox{\scriptsize +/--}$ means approaching the fault plane from the upper/lower body side, respectively). ${\B u}(x,y,t)$ for each body satisfies the Navier-Lam\'e equation $\nabla\!\cdot\!{\B \sigma}\=\frac{\mu}{1-2\nu}\nabla\!\left(\nabla\!\cdot\!{\B u}\right)+\mu\nabla^2{\B u}\=\rho\,\ddot{\B u}$, with its own shear modulus $\mu$, Poisson's ratio $\nu$ and mass density $\rho$~\cite{Landau1986}. The Cauchy stress tensor field $\B \sigma$ was related to the displacement field $\B u$ through Hooke's law and each superimposed dot represents a partial time derivative.

The fault at $y\=0$ is assumed to be described by the rate-and-state constitutive relation $\tau\=\sigma_{xy}\=-f(v,\phi)\sigma_{yy}$. Fault opening or interpenetration are excluded, i.e.~we assume $u_y(x,y\!=\!0^+,t)\=u_y(x,y\!=\!0^-,t)$, and $\sigma_{xy}$ and $\sigma_{yy}$ are continuous across the fault. The internal state field $\phi(x,t)$ evolves according to $\dot{\phi}\=g(\phi\,\dot{u}/D)$, with $g(1)\=0$ and $g'(1)\!<\!0$, as in Sect.~\ref{sec:conventional}. We then introduce interfacial slip perturbations of the form $\delta\epsilon\!\propto\!\exp(\Lambda t-i k x)$, where $\Lambda$ is the complex growth rate and $k$ is the wavenumber. The shear and normal stress perturbations are related to $\delta\epsilon$ using the solution of the quasi-static Navier-Lam\'e equation, and take the form $\delta\sigma_{xy}\=-\mu\,k\,G_1\,\delta\epsilon$, $\delta\sigma_{yy}\=i\mu\,k\,G_2\,\delta\epsilon$, $\mu$ is the shear modulus of the upper body. We focus on the quasi-static regime, i.e.~excluding inertia, because nucleation generically takes place in this regime. The quasi-static elastic transfer functions $G_1$ and $G_2$, see Supporting Information~\citep{Geubelle1995}, contain all of the information about the system's geometry, the elastic properties of the sliding bodies and loading conditions (e.g.~velocity vs.~stress boundary condition). The perturbation in the frictional resistance takes the form $\delta{f}\=\tfrac{\Lambda(a\Lambda\ell-\zeta V)}{V(V+\Lambda\ell)}\,\delta\epsilon$, where we used $\delta{v}\=\Lambda\delta\epsilon$, and the definitions $\ell\!\equiv\!-\tfrac{D}{g'(1)}\!>\!0$, $a\!\equiv\!v\tfrac{\partial\!f(v,\phi)}{\partial v}\!>\!0$ and $\zeta\!\equiv\!-v\tfrac{df(v, D/v)}{dv}\=-\tfrac{df(v,D/v)}{d\!\log{v}}$ (the latter two are evaluated at $v\=V$), see Supporting Information. Note that $\zeta$ can be both positive (velocity-weakening friction) and negative (velocity-strengthening friction) depending on the materials, the sliding velocity $V$ and physical conditions (e.g.~temperature)~\citep{Bar-Sinai2014}. For the small slip velocities regime of interest here we assume that friction is velocity-weakening, hence we consider $\zeta\!>\!0$.

The linear perturbation spectrum $\Lambda(k)$ is determined by the perturbation of the constitutive relation, which reads
\begin{equation}
\delta\tau=\delta\sigma_{xy}=\sigma_0\delta{f}-f\delta\sigma_{yy}\ .
\end{equation}
Substituting the results for $\delta\sigma_{xy}$, $\delta\sigma_{yy}$ and $\delta{f}$, we obtain an equation for $\Lambda(k)$
\begin{equation}
\label{eq:spectrum}
\mu\,k\left(G_1-i f G_2\right)+\sigma _0\frac{\Lambda(a \Lambda  \ell -\zeta  V)}{V (V+\Lambda  \ell )}=0\ .
\end{equation}
Once solutions $\Lambda(k)$ are obtained, instability is implied whenever $\Re[\Lambda(k)]\!>\!0$, corresponding to an exponential growth of perturbations. Consequently, $k_c$ is determined as the largest wavenumber $k$ (smallest wavelength) for which $\Re[\Lambda(k)]\=0$ and the critical nucleation length is estimated as $L_c\!\equiv\!2\pi/k_c$.

Solutions to Eq.~\eqref{eq:spectrum} for some cases are available in the literature. Most notably, for two identical half-spaces we have $G_1\=\text{sign}(k)[2(1-\nu)]^{-1}$ and $G_2\=0$ (see Supporting Information), where the latter represents the absence of a bimaterial effect for elastically identical materials of the same shape/geometry. Plugging these transfer functions into Eq.~\eqref{eq:spectrum}, one can readily obtain a known result for the critical wavenumber~\citep{Rice1983}, which reads $k_c\=2(1-\nu)\zeta \sigma _0 \mu^{-1} \ell^{-1}$. Using our proposed criterion $L_c\!\equiv\!2\pi/k_c$, we obtain
\begin{equation}
\label{eq:half_spaces}
L_c=\frac{\pi\,\mu\,\ell }{\zeta  (1-\nu ) \sigma _0}\qquad\quad\Longrightarrow\qquad\quad \eta=\frac{\pi }{1-\nu} \ ,
\end{equation}
where $\eta$ was defined in Eq.~\eqref{eq:1DOF_Lc}. This prediction for the critical nucleation length is identical to the one in Eq.~\eqref{eq:1DOF_Lc}, which basically follows from dimensional considerations, once the pre-factor $\eta\=\pi(1-\nu)^{-1}$ is identified as done above (and the definitions of $\ell$ and $\zeta$ are recalled). This value of the pre-factor $\eta$ is $\pi$ times larger than the largest value we have been able to trace in the available literature based on the conventional approach, hence we conclude that the proposed approach predicts a significantly larger nucleation length $L_c$ for identical half-spaces as compared to the conventional approach. Indeed, some numerical simulations of earthquake nucleation indicated that the conventional prediction with $\eta\!\simeq\!1$ quite significantly underestimates the observed $L_c$~\citep{Lapusta2003}.

The physical picture of nucleation developed in this section suggests that the {\em origin} of nucleation is a linear frictional instability, while the {\em outcome} of nucleation is typically strongly nonlinear. In particular, the critical
nucleation conditions coincide with the onset of linear instability when the patch size reaches $L_c$, then the slip velocity increases exponentially in the linear regime until nonlinearities set in when the slip velocity is large enough. Finally, the patch breaks up into propagating rupture fronts. The linear stage of the instability is expected to be rather generic, and in particular nearly independent of the exact functional form of $g(\cdot)$ (with $g(1)\=0$ and $g'(1)\!<\!0$) within the rate-and-state constitutive framework and of the background strength of the fault quantified by the initial age $\phi(t\=0)$, while the nonlinear stages that follow may depend on the details of the constitutive relation and the background fault strength.

These generic properties of the onset of nucleation will be explicitly demonstrated in Sect.~\ref{sec:FiniteH} below. Furthermore, we note that the works of~\citet{Rubin2005, Ampuero2008} apparently focus on the nonlinear stages of nucleation, which is consistent with the fact that they find differences between different friction laws and that their patches can shrink/expand during the nonlinear evolution of the instability. The nonlinear stages -- on the route to rupture propagation -- cannot take place, though, if the patch does not reach first the size $L_c$ determined by the linear instability. Hence, we believe that the above defined $L_c$ is the relevant nucleation length, and not any other length that might characterize the nonlinear evolution of the instability.

\section{Application to bimaterial interfaces}
\label{sec:bimaterial}

The general framework laid down in the previous section, unlike the conventional approach, can be naturally applied to bimaterial interfaces. We consider then two half-spaces made of different elastic materials, the upper half-space is characterized by a shear modulus $\mu\1$ and Poisson's ratio $\nu\1$ and the lower half-space by a shear modulus $\mu\2$ and Poisson's ratio $\nu\2$. It corresponds to Fig.~\ref{fig:sys_fig} (left), once the limits $H\1\!\to\!\infty$ and $H\2\!\to\!\infty$ are taken. Defining $\psi\!\equiv\!\mu\2\!/\mu\1$ and $\mu\!\equiv\!\mu\1$ (i.e.~the shear modulus of the upper body is denoted by $\mu$, as before), the elastic transfer functions for this bimaterial system take the form~\citep{Rice2001} (see also Supporting Information)
\begin{equation}
\label{eq:Gs_bimaterial}
G_1=\frac{{\C M}}{2 \mu}\text{sign}(k),\qquad\qquad G_2=\frac{\beta {\C M}}{2\mu} \ ,
\end{equation}
where
\begin{equation}
\hspace{-0.14cm}{\C M}\!\equiv\!\frac{2\psi\mu(1\!-\!\beta^2)\!^{-1}}{\psi(1\!-\!\nu\1\!)\!+\!(1\!-\!\nu\2\!)},\qquad\qquad
\beta\!\equiv\!\frac{\psi(1\!-\!2\nu\1\!)\!-\!(1\!-\!2\nu\2\!)}{2[\psi(1\!-\!\nu\1\!)\!+\!(1\!-\!\nu\2\!)]}.
\end{equation}
${\C M}$ plays the role of an effective bimaterial modulus, which approaches $\mu/(1-\nu)$ in the identical materials limit, $\mu\1\=\mu\2\=\mu$ and $\nu\1\=\nu\2\=\nu$. $\beta$, which appears in $G_2$ but not in $G_1$, vanishes in the identical materials limit (and consequently $G_2$ vanishes in this limit as well) and hence it quantifies the bimaterial effect.
\begin{figure*}[ht]
  \centering
  \includegraphics[width=0.8\textwidth]{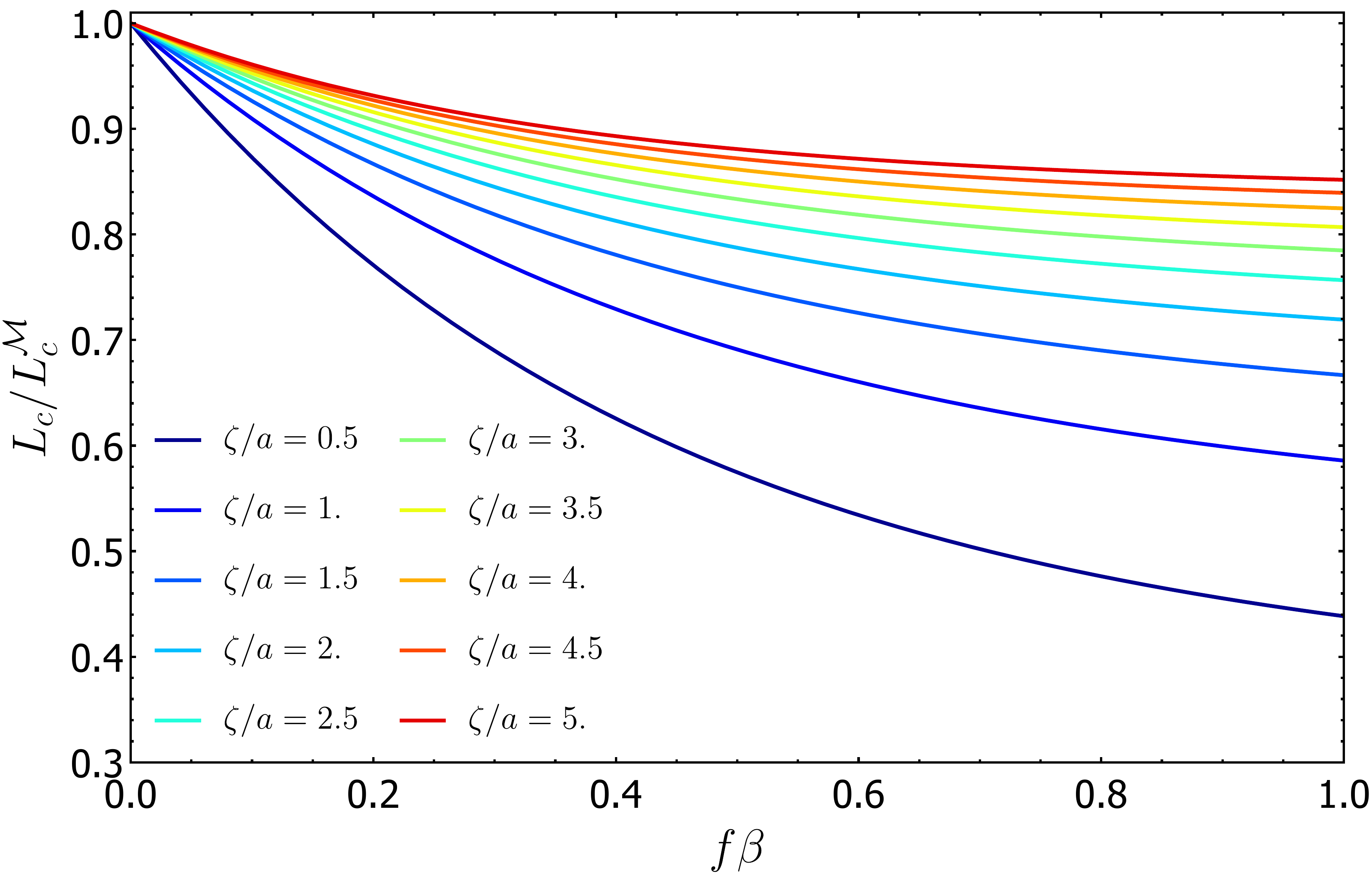}\\
  \caption{The critical nucleation length $L_c$ (normalized by $L_c^{\cal M}$) for bimaterial interfaces separating two half-spaces, cf.~Eq.~\eqref{eq:bimaterial_Lc}, plotted as a function of $f\beta$ for various values of $\zeta/a$.}
  \label{fig:bi}
\end{figure*}

The presence of a bimaterial contrast, $\beta\!\ne\!0$, introduces a new destabilization effect associated with a coupling between slip and normal stress perturbations, in addition to the the destabilizing effect associated with velocity-weakening friction, $\zeta\!>\!0$. Hence, on physical grounds one expects $L_c$ to decrease with increasing bimaterial contrast. To test this, we insert $G_{1,2}$ of Eq.~\eqref{eq:Gs_bimaterial} into Eq.~\eqref{eq:spectrum} and calculate $k_c$, obtaining the following expression for $L_c\=2\pi/k_c$
\begin{equation}
\label{eq:bimaterial_Lc}
L_c\!=\!\frac{\pi {\C M} \ell}{\zeta \sigma_0}\frac{(f\beta)^2\!\left(1+\zeta/a-\sqrt{\left(1+\zeta/a\right)^2 + \frac{\displaystyle 4\,\zeta/a}{\displaystyle (f\beta)^2}} \right)+2\,\zeta/a}{2\,\zeta/a } \ .
\end{equation}
The first multiplicative contribution on the right-hand-side, $L_c^{\cal M}\!\equiv\!\frac{\pi {\C M} \ell }{\zeta  \sigma _0}$, is obtained by replacing $\mu/(1\!-\!\nu)$ in our result in Eq.~\eqref{eq:half_spaces} by the effective modulus ${\C M}$. A similar replacement has been proposed by~\citet{Rubin2007} in the context of a different heuristic estimate of the critical nucleation length for bimaterial interfaces. Consequently, we plot in Fig.~\ref{fig:bi} $L_c$ of Eq.~\eqref{eq:bimaterial_Lc}, normalized by $L_c^{\cal M}$, as a function of $f\beta$ for various values of $\zeta/a$. It is observed that $L_c$ for bimaterial interfaces is generically {\em smaller} than the conventional estimate $L_c^{\cal M}$, indicating that bimaterial interfaces may be more unstable than previously considered. We note in passing that Eq.~\eqref{eq:bimaterial_Lc} remains valid also in the presence of velocity-strengthening friction, $\zeta\!<\!0$, for which it predicts that for sufficiently strong bimaterial contrasts, $f\beta\!\ge\!\frac{2\sqrt{-a \zeta }}{a+\zeta}$, instability is implied even for velocity-strengthening friction~\citep{Rice2001}.

\section{Application to Finite-size systems and comparison to inertial Finite-Element-Method calculations}
\label{sec:FiniteH}

The general framework laid down in section~\ref{sec:LSA}, unlike the conventional approach, can be naturally applied to finite-size systems. To demonstrate this, we consider here a system that features both finite dimensions and a bimaterial contrast. In particular, we consider a long deformable body of height $H$, and of elastic constants $\mu$ and $\nu$, in rate-and-state frictional contact with a rigid substrate under the application of a compressive stress $\sigma_0$ and a shear stress $\tau_0$. This configuration corresponds to Fig.~\ref{fig:sys_fig} (left), once the limit $\mu\2\!\to\!\infty$ is taken. In this case, the elastic transfer functions appearing in Eq.~\eqref{eq:spectrum} take the form (see Supporting Information)
\begin{equation}
\begin{split}
&G_1=\frac{4 (1-\nu ) (2 H k+\sinh (2 H k))}{2 H^2 k^2+(3-4 \nu ) \cosh (2 H k)-4 \nu  (3-2 \nu )+5}\ ,\\ &G_2=\frac{4 \left(H^2 k^2+(1-2 \nu ) \sinh ^2(H k)\right)}{2 H^2 k^2+(3-4 \nu ) \cosh (2 H k)-4 \nu  (3-2 \nu )+5} \ .
\end{split}
\label{eq:Gs_finiteH}
\end{equation}
When substituted in Eq.~\eqref{eq:spectrum}, we obtain a complex equation which is not analytically tractable, but rather is amenable to numerical analysis. Let us denote the solution by $k_c(H)$ and the corresponding prediction for the critical nucleation length by $L_c(H)\=2\pi/k_c(H)$.

\begin{figure*}[ht]
  \centering
  \includegraphics[width=0.8\textwidth]{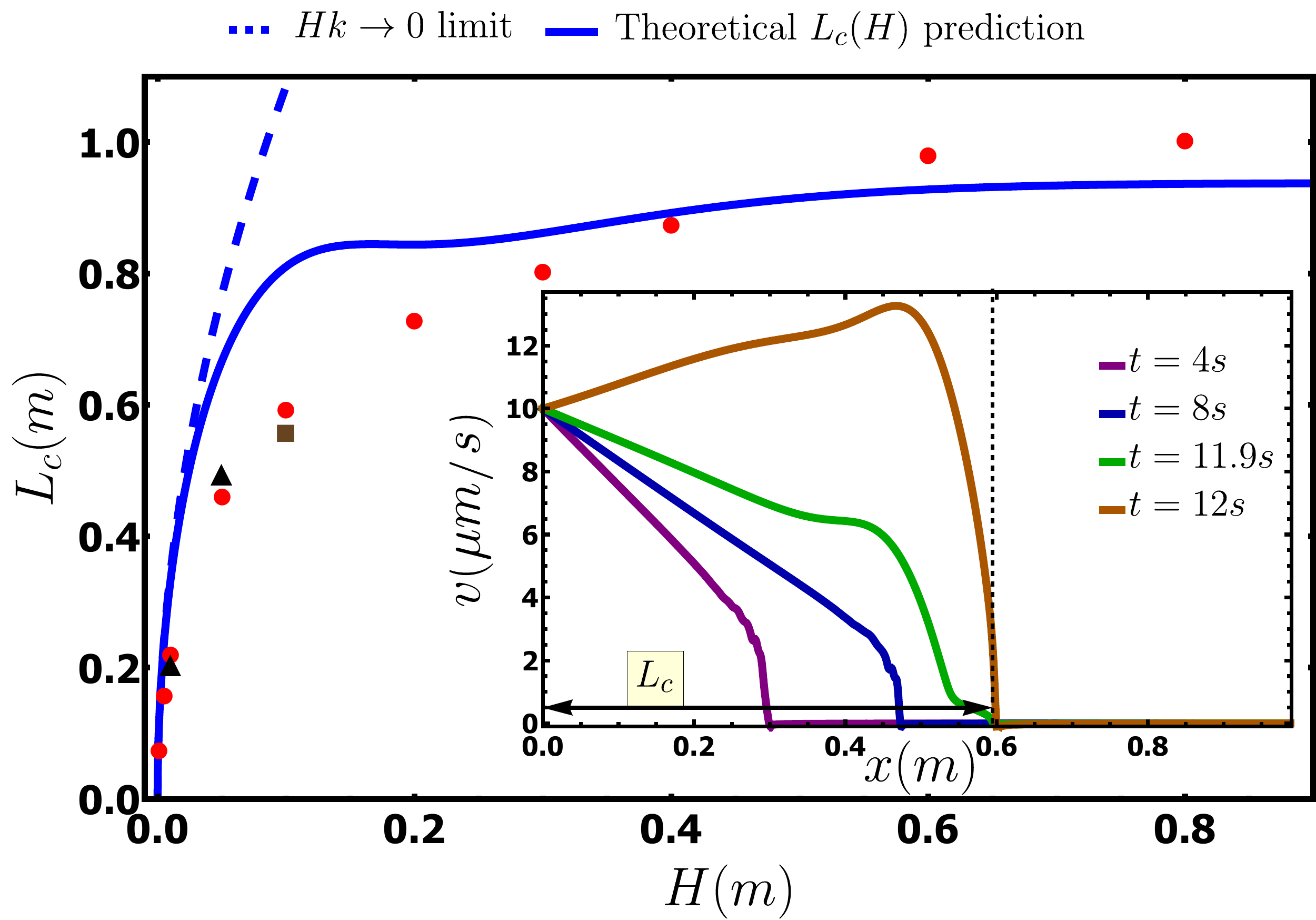}\\
  \caption{The theoretical prediction for the critical nucleation length $L_c$ for a generic rate-and-state constitutive relation as a function of the height $H$ of an elastic body sliding on top of a rigid substrate (solid line). The material, interfacial and loading parameters are given in the text. The analytic approximation for $L_c(H)$ in the $Hk\!\to\!0$ limit, cf.~Eq.~\eqref{eq:Lc1D}, is added (dashed line). The nucleation length measured in inertial FEM simulations of a finite elastic body of height $H$ under sideway loading for the aging law (see text and Fig.~\ref{fig:sys_fig} (right) for details) is shown as a function of $H$ (red circles). For $H\!=\!0.01$ m and $H\!=\!0.05$ m, $L_c$ for the slip law is also shown (black triangles), demonstrating small variation compared to the result for the aging law. For $H\!=\!0.1$ m, $L_c$ for $\phi(t\!=\!0)\!=\!10^3$ s is also shown (brown square), demonstrating small variation compared to the result for $\phi(t\!=\!0)\!=\!1$ s (i.e.~three orders of magnitude difference in the initial age of the fault). (inset) A sequence of snapshots in time (see legend) of the slip velocity field in inertial FEM simulations for the aging law with $H\!=\!0.1$ m, demonstrating the propagation of a creep patch from the loading edge at $x\!=\!0$ into the interface. At a certain creep patch length (denoted by a vertical dashed line and a horizontal double-head arrow) an instability accompanied by accelerated slip takes place. This is the numerically extracted nucleation length for this height $H$, as can be seen in the main panel.}
  \label{fig:Lc_FH}
\end{figure*}

Equation~\eqref{eq:spectrum}, with $G_{1,2}$ of Eq.~\eqref{eq:Gs_finiteH}, does admit an analytic solution in the limit $Hk\!\to\!0$, i.e.~when the system height $H$ is small compared to field variations parallel to the interface characterized by a lengthscale $\sim k^{-1}$. In this limit, we find $G_1\!\simeq\!2 H k(1\!-\!\nu)^{-1}$ and $G_2\!\simeq\!0$. Using these in Eq.~\eqref{eq:spectrum}, we obtain
\begin{equation}
L_c^{(Hk \to 0)} \simeq 2\pi \sqrt{\frac{2H \mu\,\ell }{\zeta  (1-\nu ) \sigma _0}}\ .
\label{eq:Lc1D}
\end{equation}
$L_c^{(Hk \to 0)}$ predicts the small $H$ behavior of $L_c(H)$ and constrains any numerical calculation of $L_c(H)$ to be quantitatively consistent with it in this limit. In addition, it is fully consistent with the results of~\citet{Bar-Sinai2013}. We numerically calculated $L_c(H)$ for the following set of parameters: $\mu\=3.1$ GPa, $\nu\=1/3$, $f\=0.41$, $a\=0.0068$, $\zeta\=0.016$, $\sigma_0\=1$ MPa, $\ell\=0.5\,\mu$m, and $V\=10\,\mu$m/s (the latter corresponds to an applied shear stress $\tau_0\=f(V,\phi\=D/V)\sigma_0$). The result is plotted in the main panel of Fig.~\ref{fig:Lc_FH} (solid line). When $L_c^{(Hk \to 0)}$ of Eq.~\eqref{eq:Lc1D} is superimposed on it (dashed line), perfect agreement at small $H$ and significant deviations at larger $H$ are observed, as expected.

Our goal now is to quantitatively test the ability of the calculated $L_c(H)$ to predict the critical nucleation length in a realistic situation in which a slowly expanding creep patch spontaneously nucleates accelerating slip. We would also like to test the theoretical prediction that $L_c$ is nearly independent of the specific friction law (in particular the aging vs.~the slip $\phi$ evolution laws) and the background fault strength (the initial value of $\phi$). To these aims, we performed inertial Finite-Element-Method (FEM) calculations that are directly relevant for the geometrical configuration and material parameters discussed in the last two paragraphs. In particular, we consider a deformable body of height $H$ which is also of finite extent in the direction parallel to the interface and which is loaded (by an imposed velocity $V\=10\,\mu$m/s that is initiated at $t\=0$) at its lateral edge (defined as $x\=0$), rather than at its top edge at $y\=H$, see Fig.~\ref{fig:sys_fig} (right). The advantage of this sideway loading configuration is that it naturally generates a creep patch that slowly expands from $x\=0$ along the interface, cf.~the inset of Fig.~\ref{fig:Lc_FH}. The interface is first described by the aging rate-and-state constitutive relation with $\dot{\phi}\!\simeq\!1\!-\!\phi v\!/\!D$ and $f(v,\phi)\!\simeq\!f_0\!+\!a\log(v\!/V)\!+\!(\zeta+a)\log(\phi V\!/\!D)$, where $f_0\=0.41$ and the other parameters are as above. The initial conditions are $v(t\=0)\=0$ and $\phi(t\=0)\=1s$. The full constitutive relation used in the FEM calculations, which also allows a transition from stick ($v\=0$) to slip ($v\!>\!0$), can be found in the supporting information~\cite{Hecht2012}.

Our theoretical approach predicts that the creep patch loses its stability and develops accelerating slip upon reaching a certain critical length. This is indeed observed in the inset of Fig.~\ref{fig:Lc_FH}, where the slip velocity blows up when the creep patch reaches a certain length. We then measured the critical length in inertial FEM calculations for different system heights $H$ (in addition to the inset of Fig.~\ref{fig:Lc_FH}, see also the supporting information for the details of the determination of $L_c$ in the numerical calculations) and superimposed the results for the aging law (red circles) on the main panel of Fig.~\ref{fig:Lc_FH}. It is observed that the theoretical prediction for the critical nucleation length $L_c(H)$ is in reasonably good quantitative agreement with the FEM results for the full range of system heights $H$. This major result lends serious support to the approach developed in this Letter.

In order to test whether the theoretically predicted critical nucleation length $L_c(H)$ is indeed nearly independent of the details of the friction law, we repeated the above described FEM calculations for $H\=0.01$ m and $H\=0.05$ m with the slip law instead of the aging law; that is, we used $\dot{\phi}\!\simeq\!-(\phi v\!/\!D)\log(\phi v\!/\!D)$ (the full constitutive relation can again be found in the supporting information). The resulting critical nucleation length (black triangles in main panel of Fig.~\ref{fig:Lc_FH}) for both $H$ values exhibits only a small variation (less than $10\%$) compared to the results for the aging law. Furthermore, we repeated the above described FEM calculations for the aging law with $H\=0.1$ m, except that we increased the initial age of the fault by three orders of magnitude, from $\phi(t\=0)\=1$ s to $\phi(t\=0)\=10^3$ s. The resulting critical nucleation length (brown square in main panel of Fig.~\ref{fig:Lc_FH}) exhibits only a small variation (less than $10\%$) compared to the result for $\phi(t\=0)\=1$ s. These results lend strong support to the idea that the critical nucleation length $L_c$ is determined by a linear instability that is reasonably predicted by the procedure developed in this Letter.

\section{Concluding remarks}
\label{sec:conclusion}

In this Letter we developed a theoretical approach for the calculation of the critical nucleation length for accelerating slip $L_c$. The proposed approach builds on existing literature by adopting the view that nucleation is associated with a linear frictional instability of an expanding creep patch. It deviates from the conventional approach in the literature by replacing the problem of the stability of a spatiotemporally varying creep patch by an effective homogeneous sliding linear stability analysis for deformable bodies, rather than invoking a spring-block stability analysis supplemented with some fracture mechanics estimates for deformable bodies. The quality of the predictions emerging from the proposed approach therefore depend on the degree by which the creep patch can be approximated by spatially homogeneous fields. This approximation is expected to be reasonable in many cases in light of the weak/logarithmic velocity dependence of friction in many materials. The temporal aspects of the creep patch propagation are taken into account by the requirement that it becomes unstable upon attaining a length for which an unstable mode from the homogeneous linear stability analysis can be first fitted into.

The proposed approach is rather general and applies to a broad range of physical situations. For sliding along rate-and-state frictional interfaces separating identical elastic half-spaces, it has been shown to predict a significantly larger nucleation length compared to the conventional approach. For sliding along rate-and-state frictional interfaces separating different elastic half-spaces, the proposed approach has been shown to predict a bimaterial weakening effect which appears to be stronger than previously hypothesized, resulting in a smaller nucleation length. Finally, the proposed approach has been applied to finite-height systems. For this case, the scenario of a loss of stability of an expanding creep patch has been directly demonstrated using inertial FEM calculations and the predicted nucleation length has been shown to be in reasonably good quantitative agreement with direct FEM results for a range of system heights. The quality of the theoretical predictions has been shown to be nearly independent of the specific friction law used (aging vs.~slip laws) and the background strength of the fault. These results offer a theoretical framework for predicting rapid slip nucleation along faults and hence may give rise to better short-term earthquake prediction capabilities. The proposed approach can and should be quantitatively tested in a wide variety of interfacial rupture nucleation problems, using both theoretical tools and extensive numerical simulations.

\begin{acknowledgments}
E.~B.~acknowledges support from the Israel Science Foundation (Grant No.~295/16), the William Z.~and Eda Bess Novick Young Scientist Fund, COST Action MP1303, and the Harold Perlman Family. R.~S.~acknowledges support from the DFG priority program 1713. We are grateful to Eric Dunham, one of the reviewers of the manuscript, for his valuable and constructive comments and suggestions. We also thank Robert Viesca for useful discussions in the context of nucleation on bimaterial interfaces. M.~A.~acknowledges Yohai Bar-Sinai for helpful guidance and assistance.  The analytical formulae and numerical methods described in the main text and supporting information are sufficient to reproduce all the results and plots presented in the paper.
\end{acknowledgments}

\onecolumngrid
\newpage
\begin{center}
	\textbf{\large Supplemental Materials for:\\``Critical nucleation length for accelerating frictional slip''}
\end{center}

\setcounter{equation}{0}
\setcounter{figure}{0}
\setcounter{section}{0}
\setcounter{table}{0}
\setcounter{page}{1}
\makeatletter
\renewcommand{\theequation}{S\arabic{equation}}
\renewcommand{\thefigure}{S\arabic{figure}}
\renewcommand{\thesection}{S-\Roman{section}}
\renewcommand*{\thepage}{S\arabic{page}}
\renewcommand{\bibnumfmt}[1]{[S#1]}
\renewcommand{\citenumfont}[1]{S#1}

\section*{Introduction}

The goal of this Supporting Information file is to provide additional details regarding the derivation of the theoretical results (Text S1) and Finite-Element-Method (FEM) calculations (Text S2) appearing in the main text.

\section*{Text S1: Derivation of theoretical results appearing in the main text}

\subsection*{1.~The transfer functions \texorpdfstring{$G_{1}$}{} and \texorpdfstring{$G_{2}$}{} in plane-strain elasticity}
\label{sec:M}

We consider a two-dimensional (2D) elastic body that occupies the region $-\infty\!<\!x\!<\!\infty$ and $0\!\le\!y\!\le\!H$ under plain-strain conditions. The bottom boundary of the body at $y\=0$ is in contact with another body and hence some boundary tractions are generated at the interface. The latter are described by the interfacial stress components  $\sigma_{yi}$ (constituting a vector where $i\=x,y$). Our first goal would be to explicitly calculate the quasi-static (i.e.~excluding inertia) relation between $\sigma_{yi}$ and the interfacial displacement $\bm u$ in the form $\delta{u}_i\=M_{ij}\delta\sigma_{yj}$, i.e.~to calculate the matrix $\bm M$. The body satisfies quasi-static equilibrium (momentum balance) and linear elasticity (Hooke's law)~\citep{Landau1986_SI}
\begin{align}
\label{eq:LE}
\nabla \cdot \bm{\sigma}&=0\ ,\qquad
&
\left(
\begin{array}{c}
 \sigma _{xx} \\
 \sigma _{yy} \\
 \sigma _{xy} \\
\end{array}
\right)&=\frac{2 \mu}{1-2 \nu } \left(
\begin{array}{ccc}
 1-\nu & \nu & 0 \\
 \nu& 1-\nu& 0 \\
 0 & 0 & 1-2 \nu \\
\end{array}
\right) \left(
\begin{array}{c}
 \varepsilon _{xx} \\
 \varepsilon _{yy} \\
 \varepsilon _{xy} \\
\end{array}
\right)\ ,
\end{align}
where $\varepsilon_{ij}\!\equiv\!\frac{1}{2}\left(\partial_i u_{j}+\partial_j u_{i}\right)$ is the infinitesimal strain tensor (not to be confused with the slip displacement discontinuity vector $\epsilon_i$), $\bm \sigma$ is Cauchy's stress tensor, and $\mu$ and $\nu$ are the shear modulus and Poisson's ratio, respectively.

At the top boundary $y\=H$ the body is loaded by imposing a horizontal velocity $V$ and a compressive normal stress $\sigma_{yy}\=-\sigma_0$ (with $\sigma_0\!>\!0$). The homogeneous solution $\bm u\h$ consistent with these boundary conditions reads
\begin{equation}
\bm u\h(y,t)\equiv \Big(V t+\frac{f\sigma_0}{\mu}y\ ,\ -\frac{1-2\nu}{2(1-\nu)}\frac{\sigma_0}{\mu} y\Big)\ .
\end{equation}
Since the equations of motion are linear, one can decompose a general solution to a sum of the steady solution of homogeneous sliding and a deviation from it, and write $\bm u(x,y,t)\=\bm u\h(y,t)+\delta\bm u(x,y,t)$. The boundary conditions (BC) at $y\=H$ are
\begin{equation}
 \pa_t(\delta u_x)=0 \qquad \mbox{ and } \qquad \delta\sigma_{yy}=0 \ .
 \label{eq:SBC}
\end{equation}

Consider now a single Fourier mode, i.e. assume that all fields depend on $x$ and $t$ as $\propto e^{\Lambda t-i k x}$, for which Eq.~\eqref{eq:LE} admits a solution of the form
\begin{equation}
\label{eq:full_sol}\begin{split}
&\delta\bm u=e^{\Lambda t-i k x}\times\\
&\left(\begin{array}{cccc}
 A_1 e^{k y}+A_2 e^{-k y}+A_3 k y e^{k y}+A_4 k y e^{-k y}\\
 i A_1 e^{k y}-i A_2 e^{-k y}+i A_3\left(k y-\kappa\right)e^{k y}-i A_4\left(k y+\kappa\right)e^{-k y}
 \end{array}\right)\ ,
\end{split}\end{equation}
where $A_i$ are $4$ (yet) unknown amplitudes which are determined by employing $4$ boundary conditions, and $\kappa\!\equiv\!3\!-\!4\nu$. Note that Eq.~\eqref{eq:full_sol} remains valid for plain-stress conditions, but with $\kappa\!=\!\frac{3-\nu}{1+\nu}$. Applying two boundary conditions at $y\=H$, cf.~Eq.~\eqref{eq:SBC}, the solution takes the form
\begin{equation}
\label{eq:sol_BC}\begin{split}
&\delta\bm u=e^{\Lambda t-i k x}\times\\
&\left(\begin{array}{cccc}
 B_1 \sinh \left(q\left(1-\tilde{y}\right)\right)-B_2 q \left(e^{-q \left(1-\tilde{y}\right)}-\tilde{y} \cosh \left(q\left(1-\tilde{y}\right)\right)\right)\\
 -iB_1 \cosh \left(q\left(1-\tilde{y}\right)\right)-iB_2 \left(\kappa\cosh \left(q\left(1-\tilde{y}\right)\right)+q\left(e^{-q \left(1-\tilde{y}\right)}+\tilde{y} \sinh \left(q\left(1-\tilde{y}\right)\right)\right)\right)
 \end{array}\right),
\end{split}\end{equation}
where $q\!\equiv\!Hk$, $\tilde{y}\equiv\!y/H$ and the two amplitudes $B_1$ and $B_2$ remain unspecified (they depend on the contact interactions at the interface, which remain unspecified). Using Hooke's law and momentum balance, cf.~Eq.~\eqref{eq:LE}, one can express the relation between the interfacial displacements $u_i\left(\tilde{y}\!=\!0\right)$ and the interfacial stresses $\sigma_{yi}\left(\tilde{y}\!=\!0\right)$, for any $B_1$ and $B_2$, in the form $\delta{u}_i\=M_{ij}(k)\delta\sigma_{yj}$~\citep{Geubelle1995_SI}, where
\begin{equation}
\label{eq:general_M}\begin{split}
\bm{M}=&\frac{1}{2k\mu\left(\sinh\left(2 H k\right)-2 H k\right)}\times\\
&\left(
\begin{array}{cc}
 -4 (1-\nu ) \sinh ^2(H k) & i (2 H k+(1-2 \nu ) \sinh (2 H k)) \\
 -i (2 H k+(1-2 \nu ) \sinh (2 H k)) & -4 (1-\nu ) \cosh ^2(H k) \\
\end{array}
\right)
\ .
\end{split}\end{equation}
Note that in case that the body under consideration occupies the region in space $-H\!\le\!y\!\le\!0$ (with $H\!>\!0$) the analysis remains valid, but $H$ should be replaced by $-H$. This simply amounts to changing the sign of the diagonal entries of $\bm M$ in Eq.~\eqref{eq:general_M}. In the limit $Hk\!\to\!\infty$, the matrix $\bm M$ takes the form
\begin{equation}\label{eq:M_inf}
\bm{M}=\frac{1}{2k\mu}
\left(
\begin{array}{cc}
 -2 (1-\nu) \text{sign}(Hk) & i (1-2 \nu ) \\
 -i (1-2 \nu) & -2 (1-\nu) \text{sign}(Hk) \\
\end{array}
\right)
\ .
\end{equation}

The matrix $\bm M$ characterizes a single body. Next, we aim at calculating the response of a general composite system, composed of two bodies made of different linear elastic materials of different heights in frictional contact (both bodies are assumed to be infinite in the $x$-direction). The upper body, denoted by the superscript (1), is assumed to occupy the region $0\!<\!y\!<H\1$ and the lower one, denoted by the superscript (2), the region $-H\2\!<\!y\!<\!0$ (with positive $H^{(i)}$). Since $\mu$ and $\nu$ may be different for the different bodies, they are also labeled with a superscript. Following the previous derivation, the displacements at the frictional interface $y\=0$ are given as
\begin{align}
 \begin{pmatrix}\delta{u}_x\1\\\delta{u}_{y}\1\\\end{pmatrix}&=\bm M\1 \begin{pmatrix}\delta\sigma_{yx}\1\\\delta\sigma_{yy}\1\\\end{pmatrix}\ ,
 &
 \begin{pmatrix}\delta{u}_x\2\\\delta{u}_{y}\2\\\end{pmatrix}&=\bm M\2 \begin{pmatrix}\delta\sigma_{yx}\2\\\delta\sigma_{yy}\2\\\end{pmatrix}\ .
\end{align}
Since both $\sigma_{yx}\=\sigma_{xy}$ and $\sigma_{yy}$ are continuous at $y\=0$, the displacement discontinuity can be written as
\begin{equation}
 \begin{pmatrix}\delta{u}_x\1-\delta{u}_x\2\\\delta{u}_y\1-\delta{u}_y\2\\\end{pmatrix}=\Big(\bm M\1 - \bm M\2 \Big)\begin{pmatrix}\delta\sigma_{yx}\\ \delta\sigma_{yy} \\\end{pmatrix}\ .
\end{equation}
Since we exclude opening gaps $u_y\1-u_y\2\!=\!0$, but allow a slip discontinuity $\delta\epsilon\!\equiv\!\delta{u}_x\1-\delta{u}_x\2$  (cf.~the main text), we obtain the following relation between $\delta\sigma_{yi}$ and $\delta\epsilon$~\citep{Geubelle1995_SI}
\begin{align}
\begin{pmatrix}\delta\sigma_{yx}\\ \delta\sigma_{yy} \\\end{pmatrix}&=\bm G \begin{pmatrix}\delta\epsilon \\ 0 \end{pmatrix}\ ,
\qquad \qquad
 \mbox{ where } \qquad
 \bm G\equiv \Big(\bm M\1 - \bm M\2 \Big)^{-1} \ .
 \label{eq:defG}
 \end{align}
 We thus write
\begin{align}\label{eq:sig_G}
 \delta\sigma_{yx}=\delta\sigma_{xy}&=-\mu\1\,k\,G_1\,\delta\epsilon \ ,
 &
 \delta\sigma_{yy}&=i\mu\1\,k\,G_2\,\delta\epsilon \ ,
\end{align}
as in Sect.~3 in the main text, with the definitions $G_1\!\equiv\!-G_{xx}/(\mu\1 k)$ and $G_2\!\equiv\!-iG_{yx}/(\mu\1 k)$. Note that in systems with reflection symmetry with respect to the interface $\bm M\2$ is obtained from $\bm M\1$ simply by taking $H\to-H$. Therefore, the diagonal terms of $\bm M\1$ and $\bm M\2$ have opposite signs and the off-diagonal terms are identical, or in other words, $\bm M\1-\bm M\2$ is diagonal. Thus, $\bm G$ is also diagonal, i.e.~$G_2\=0$, and no coupling exists between tangential motion and normal traction in this case (i.e.~for symmetric systems)~\citep{ranjith2001_SI}. For example, in the case addressed in Sec.~3 of the main text we have $\mu\1=\mu\2\equiv\mu$, $\nu\1=\nu\2\equiv\nu$, and $H\n\!\to\!\infty$ for $n\=1,2$ these in Eqs.~\eqref{eq:M_inf} and \eqref{eq:defG}, one obtains~\citep{Rice2001_SI}
\begin{align}\label{eq:G_gen}
G_1&=\frac{\text{sign}(k)}{2 (1-\nu )} ,&
G_2&=0\ .
\end{align}

\subsection*{2.~Application to bimaterial interfaces}
\label{sec:bi}

In Sec.~4 in the main text we apply our approach to the case of two half-spaces made of different materials. In this case, one can use Eq.~\eqref{eq:M_inf} for $\bm{M}\n$. For the ease of notation we define $\mu\1\!=\!\mu $ and $\mu\2\!=\!\psi\mu$, leading to~\citep{Rice2001_SI}
\begin{equation}\label{eq:M_bi}\begin{split}
\bm{M}\1&=\frac{1}{2k\mu}
\left(
\begin{array}{cc}
 -2 (1-\nu\1) \text{sign}(k) & i (1-2 \nu\1) \\
 -i (1-2 \nu\1) & -2 (1-\nu\1) \text{sign}(k) \\
\end{array}
\right)
\ ,\\
\bm{M}\2&=\frac{1}{2k\psi\mu}
\left(
\begin{array}{cc}
 2 (1-\nu\2) \text{sign}(k) & i (1-2 \nu\2) \\
 -i (1-2 \nu\2) & 2 (1-\nu\2) \text{sign}(k) \\
\end{array}
\right)
\ .
\end{split}\end{equation}
This leads to
\begin{equation}
\label{eq:Gs_bimaterial_SI}
G_1=\frac{{\C M}}{2 \mu}\text{sign}(k)\ ,\qquad\qquad\qquad G_2=\frac{\beta {\C M}}{2\mu} \ ,
\end{equation}
where
\begin{equation}
\hspace{-0.14cm}{\C M}\!\equiv\!\frac{2\psi\mu(1\!-\!\beta^2)\!^{-1}}{\psi(1\!-\!\nu\1\!)\!+\!(1\!-\!\nu\2\!)} \ ,\qquad\qquad\qquad
\beta\!\equiv\!\frac{\psi(1\!-\!2\nu\1\!)\!-\!(1\!-\!2\nu\2\!)}{2[\psi(1\!-\!\nu\1\!)\!+\!(1\!-\!\nu\2\!)]} \ ,
\end{equation}
which identify with Eqs.~(7)-(8) in the main text.

\subsection*{3.~Application to finite height systems}
\label{sec:H}

In Sec.~5 in the main text a finite height $H$ body in frictional contact with an infinitely rigid substrate was considered. The requirement that the bottom layer is infinitely rigid leads to the boundary condition $u_y\!=\!0$ at the interface. In addition, since we consider here a constant shear stress at the top boundary, rather than a constant tangential velocity, the BC's in Eq.~\eqref{eq:SBC} are replaced by
\begin{equation}
 \delta\sigma_{yx}(y\!=\!H)=0 \ , \qquad \qquad \delta\sigma_{yy}(y\!=\!H)=0\qquad \mbox{ and } \qquad u_y(y\!=\!0)=0\ .
 \label{eq:BC_rig}
\end{equation}

Using Eq.~\eqref{eq:BC_rig}, we obtain a modified solution
\begin{equation}
\label{eq:sol_inf}\begin{split}
&\delta\bm u=Ae^{\Lambda t-i k x}\times\\
&\left(\begin{array}{cccc}
\frac{2 \left(2 q^2 \left(\tilde{y}-1\right) \cosh \left(q \tilde{y}\right)+q \left((4 \nu -3) \left(\tilde{y}-2\right) \sinh \left(q \tilde{y}\right)-\tilde{y} \sinh \left(q \left(\tilde{y}-2\right)\right)\right)+(4 \nu -3) \cosh \left(q \left(\tilde{y}-2\right)\right)+(4 (3-2 \nu ) \nu -5) \cosh \left(q \tilde{y}\right)\right)}{2q+e^{2 q}+3-4 \nu}\\
 i\frac{4 \left(q^2 \tilde{y}+4 \nu ^2-6 \nu -q^2+2\right) \sinh \left(q \tilde{y}\right)-2 q \tilde{y} \left((3-4 \nu ) \cosh \left(q \tilde{y}\right)+\cosh \left(q \left(\tilde{y}-2\right)\right)\right)}{2 q+e^{2 q}+3-4\nu}
 \end{array}\right),
\end{split}\end{equation}
with one undetermined amplitude $A$. Since in this case $\delta\epsilon\!=\!\delta u_x$, $G_1$ and $G_2$ can be calculated directly from Eq.~\eqref{eq:sol_inf}, without any matrix algebra. The solution reads
\begin{equation}
\begin{split}
&G_1=\frac{\displaystyle 4 (1-\nu ) (2 H k+\sinh (2 H k))}{\displaystyle 2 H^2 k^2+(3-4 \nu ) \cosh (2 H k)-4 \nu  (3-2 \nu )+5}\ ,\\ &G_2=\frac{\displaystyle 4 \left(H^2 k^2+(1-2 \nu ) \sinh ^2(H k)\right)}{\displaystyle 2 H^2 k^2+(3-4 \nu ) \cosh (2 H k)-4 \nu  (3-2 \nu )+5} \ ,
\end{split}
\end{equation}
as appears in Eq.~(10) in the main text.

\subsection*{4.~The linear stability spectrum and finding the critical wavenumber}

The transfer functions $G_1$ and $G_2$ are plugged in Eq.~(4) in the main text. The latter transforms into a linear stability spectrum equation once $\delta{f}$ is calculated, which is done as follows~\citep{Ruina1983_SI,Rice1983_SI}. As $f$ is a function of $v$ and $\phi$, we have
\begin{equation}
\delta{f}=\frac{\partial\!f(v,\phi)}{\partial v}\delta{v}+\frac{\partial\!f(v,\phi)}{\partial\phi}\delta\phi=
\left(\frac{\partial\!f(v,\phi)}{\partial v}+\frac{\partial\!f(v,\phi)}{\partial\phi}\frac{\delta\phi}{\delta{v}}\right)\delta{v}\ ,
\end{equation}
where all the derivatives are evaluated at $v\=V$ and $\phi\=D/V$. Next we insert $v\!=\!V+\varepsilon e^{\Lambda t-i k x}$ and $\phi\!=\!\tfrac{D}{V}+\varepsilon{A}_\phi e^{\Lambda t-i k x}$ into the $\phi$ evolution equation $\dot\phi\!=\!g\left(\tfrac{v\phi}{D}\right)$, with $g(1)\!=\!0$ and $g'(1)<0$, and expand to leading order in $\varepsilon$ to obtain
\begin{equation}
\label{eq:g}
\varepsilon  e^{\Lambda  t-i k x}\Lambda  A_{\phi }=\varepsilon  e^{\Lambda  t-i k x}\frac{g'(1) \left(V^2 A_{\phi }+D\right)}{D V}\ \qquad\Longrightarrow\qquad
A_\phi=\frac{D g'(1)}{V \left(D \Lambda -V g'(1)\right)}=\frac{\delta\phi}{\delta{v}}\ .
\end{equation}
Using $\partial_V f\left(V,\frac{D}{V}\right)\=\partial_vf\left(v,\phi\right)-\frac{D}{V^2}\partial_\phi f\left(v,\phi\right)$ (evaluated at steady state, as described above), together with $\delta{v}\=\Lambda\delta\epsilon$ and Eq.~\eqref{eq:g}, we find
\begin{equation}\label{eq:delta_f}\begin{split}
\delta{f}=&\Lambda\left(\partial_vf\left(v,\phi\right)+\frac{V^2}{D}\left(\partial_vf\left(v,\phi\right)-\partial_V f\left(V,\frac{D}{V}\right)\right)\frac{D g'(1)}{V \left(D \Lambda -V g'(1)\right)}\right)\delta\epsilon=\\
&\Lambda\left(\frac{D \Lambda\partial_vf\left(v,\phi\right)- V g'(1)\partial_V f\left(V,\frac{D}{V}\right)}{D \Lambda -V g'(1)}\right)\delta\epsilon=
\Lambda\frac{a \Lambda  \ell -\zeta  V}{V (V+\Lambda  \ell )}\delta\epsilon\ ,
\end{split}\end{equation}
where we have defined $\ell\!\equiv\!-\tfrac{D}{g'(1)}\!>\!0$, $a\!\equiv\!v\tfrac{\partial\!f(v,\phi)}{\partial v}\!>\!0$ and $\zeta\!\equiv\!-v\tfrac{df(v, D/v)}{dv}\=-\tfrac{df(v,D/v)}{d\!\log{v}}$ (the latter two are evaluated at $v\=V$). Substituting $\delta{f}$ of Eq.~\eqref{eq:delta_f} into Eq.~(4) in the main text, we obtain the general linear stability spectrum in Eq.~(5) in the main text, which is reproduced here
\begin{equation}
\label{eq:spectrum_SI}
S(\Lambda,k)=\mu\,k\left(G_1-i f G_2\right)+\sigma _0\frac{\Lambda(a \Lambda  \ell -\zeta  V)}{V (V+\Lambda  \ell )}=0\ ,
\end{equation}
with the extra notation $S(\Lambda,k)$ which is omitted in the main text. Equation~\eqref{eq:spectrum_SI} is then used to find the critical wavenumber. To that aim, we look for solutions to Eq.~\eqref{eq:spectrum_SI} of the form $S(i\omega,k_c)\!=\!0$, where $\omega$ is a real frequency~\citep{Rice1983_SI}. Inserting this into Eq.~\eqref{eq:spectrum_SI} and using the fact that all the quantities in the equation are real now, we split the equation into its real and imaginary part as follows
\begin{equation}
\label{eq:ReIm}
\mu  k_c G_1 =\sigma _0\frac{\omega ^2 \ell  (a+\zeta )}{V^2+\omega ^2 \ell ^2}\ ,\qquad\qquad\qquad
f G_2 \mu  k_c=\sigma _0\frac{\omega  \left(a \omega ^2 \ell ^2-\zeta  V^2\right)}{V^3+V \omega ^2 \ell ^2}\ .
\end{equation}
These equations are solved, either analytically (as in Sec.~4 in the main text) or numerically (as in Sec.~5 in the main text). Notice that if $G_2\!=\!0$, i.e.~in the absence of coupling between slip and normal stress variations, Eq.~\eqref{eq:ReIm} implies that $\omega\!=\!\sqrt{\frac{\zeta}{a }}\frac{ V}{\ell }$, which is independent of material parameters or external normal loading. Once Eq.~\eqref{eq:ReIm} is solved, we obtain our prediction for the critical nucleation length $L_c\!=\!2\pi/k_c$. If more than one solution exists, we choose the largest value of $k_c$, as we are interested in a minimal nucleation length.

\section*{Text S2: Details of the Finite-Element-Method (FEM) calculations}

We consider a deformable body of height $H$ which is also of finite extent $L$ in the direction parallel to the contact interface with a rigid substrate (defined by $y\=0$). $L$ is taken to be much larger than the nucleation length, $L\!\gg\!L_c$, such that it plays no role in the obtained results. The deformable body is loaded by an imposed velocity $V\=10\,\mu$m/s at its lateral edge (defined as $x\=0$), initiated at $t\=0$. The top boundary is under compressive stress of magnitude $\sigma_0\=1$ MPa and no other forces are externally applied. The interface is initially at rest, $v\=0$, and its ``age'' is set to $\phi\=1$ s. The complete set of boundary conditions for the displacement vector field ${\bm u}(x,y,t)$ and the stress tensor field ${\bm \sigma}(x,y,t)$ take the form (for $t\!\ge\!0$)
\begin{align}
&u_x(x\!=\!0,y,t)\!=\!V t, \quad \sigma_{xy}(x\!=\!0,y,t)\!=\!0, \quad \sigma_{xx}(x\!=\!L,y,t)\!=\!0, \quad \sigma_{xy}(x\!=\!L,y,t)\!=\!0,  \nonumber\\
&\sigma_{yy}(x,y\!=\!H,t)\!=\!-\sigma_0,  \quad \sigma_{xy}(x,y\!=\!H,t)\!=\!0, \quad u_y(x,y\!=\!0,t)\!=\!0, \nonumber\\ &\sigma_{xy}(x,y\!=\!0,t)\!=\!-f(v,\phi)\,\sigma_{yy}(x,y\!=\!0,t),
\end{align}
where $f(v,\phi)$ is the friction law.

\subsection*{1.~The friction law}

\begin{figure}[h]
  \centering
  \includegraphics[width=0.8\textwidth]{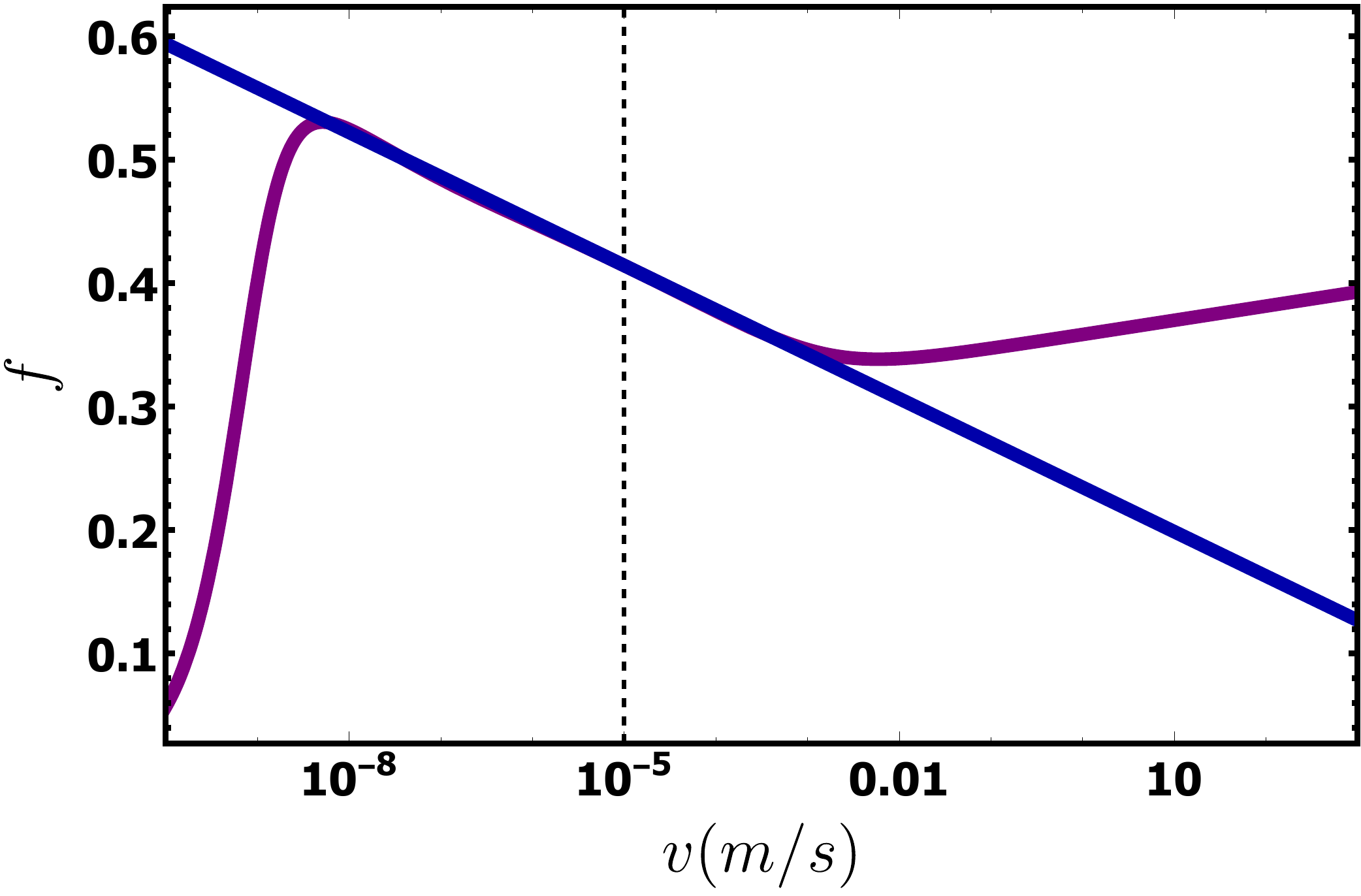}
  \caption{The steady state friction curve (purple line) corresponding to the full model in Eqs.~\eqref{eq:f}-\eqref{eq:phi} and the approximated one (blue line) corresponding to Eqs.~\eqref{eq:log}-\eqref{eq:phi_approx}. The vertical dashed line corresponds to $V$.}\label{fig:fss}
\end{figure}

The friction law used in the Finite-Element-Method (FEM) calculations should be fully consistent with the aging rate-and-state friction law presented in the main text, but must also go beyond it. The reason for this is that conventional rate-and-state models do not describe the transition from stick ($v\=0$) to slip ($v\!>\!0$), which is essential for spatiotemporal nucleation dynamics in general and for the propagation of a creep patch in particular. In fact, they feature a divergence in the $v\!\to\!0$ limit. To provide a physically sensible description of the transition from stick to slip, and to regularize the friction law, we used
\begin{equation}
\label{eq:f}
f\left(v,\phi\right)=\left(1+b \log \left(1+\frac{\phi }{\phi^*}\right)\right) \left(\frac{\theta}{\sqrt{1+\left(v_0/v\right)^2}}+\xi\log \left(1+\frac{v}{v^*}\right)\right)
\end{equation}
and
\begin{equation}
\label{eq:phi}
\dot{\phi}=1-\frac{\phi\sqrt{v^2+v_0^2}}{D} \ ,
\end{equation}
where $\theta$ and $\xi$ are dimensionless parameters, $\phi^*$ and $v^*$ are short time and slip velocity scales, and $v_0$ is an extremely small regularization slip velocity. $b$ is the ordinary aging coefficient. The values of all parameters are given in Table~\ref{tab:values}. The existence of an extremely small, yet finite, $v_0$ in the above equations ensures that $f$ vanishes in the limit $v\!\to\!0$ and that stick conditions are properly described. The friction law of Eqs.~\eqref{eq:f}-\eqref{eq:phi}, under steady state conditions, is shown in Fig.~\ref{fig:fss} (purple line). Note that the steady state friction curve exhibits a transition to velocity-strengthening friction~\citep{Bar-Sinai2014_SI} above some characteristic slip velocity at which the curve attains a minimum, but it plays no role in the nucleation process. The latter is dominated by a smaller slip velocity $V$, well within the velocity-weakening branch of the friction law (denoted by a vertical dashed line in Fig.~\ref{fig:fss}).

While the $v_0$-regularization is essential near the leading edge of the expanding creep patch, where the transition from stick to slip takes place, for most of the creep patch --- where the typical slip velocity is $V\!\gg\!v_0$ --- it is irrelevant and the conventional rate-and-state equations presented in the main text are valid. To see this, we expand the full friction law around $V\=10^{-5}$ m/s, which is $4$ order of magnitude larger than $v_0$, to obtain
\begin{equation}
\label{eq:log}
f\left(v,\phi\right) \simeq f_0+a\log \left(\frac{v}{V}\right)+\left(\zeta+a\right)\log \left(\frac{V\phi}{D}\right)\ ,
\end{equation}
and
\begin{equation}
\label{eq:phi_approx}
\dot{\phi}\simeq 1-\frac{\phi v}{D} \ ,
\end{equation}
with $f_0\!=\!0.414$, $a\!=\!0.00682$ and $\zeta\!=\!0.0156$, as appears in the main text (the values are reported in a rounded format). Finally, the steady state friction curve corresponding to Eqs.~\eqref{eq:log}-\eqref{eq:phi_approx} is superimposed on Fig.~\ref{fig:fss} (straight blue line). It is observed that the approximated rate-and-state friction law agrees extremely well with the full friction law over a huge range of slip velocities within the velocity-weakening branch.

When the slip law is used, Eq.~\eqref{eq:phi} is replaced by
\begin{equation}
\label{eq:phi_slip}
\dot{\phi}=-\left(\frac{\phi\sqrt{v^2+v_0^2}}{D}\right)\log\left(\frac{\phi\sqrt{v^2+v_0^2}}{D}\right) \ ,
\end{equation}
which is very well approximated by $\dot{\phi}\!\simeq\!-\left(\frac{\phi v}{D}\right)\log\left(\frac{\phi v}{D}\right)$ almost everywhere in space throughout the dynamics.

\subsection*{2.~Numerical implementation}

The problem defined above is solved using the FEM software package FreeFem++~\citep{Hecht2012_SI}. The partial differential equations are solved in a monolithic scheme on a triangular mesh using a real space implementation. The fully inertial evolution equations are expressed in weak form in the spirit of the finite element method. The time derivatives are expressed via finite differences. The frictional boundary condition is implemented via a semi-implicit integration scheme. Prior to the initiation of the sideway loading, the elastic fields corresponding to the uniform compressive stress, with no slip at the interface, are computed.

\begin{figure}[h]
  \centering
  \includegraphics[width=0.8\textwidth]{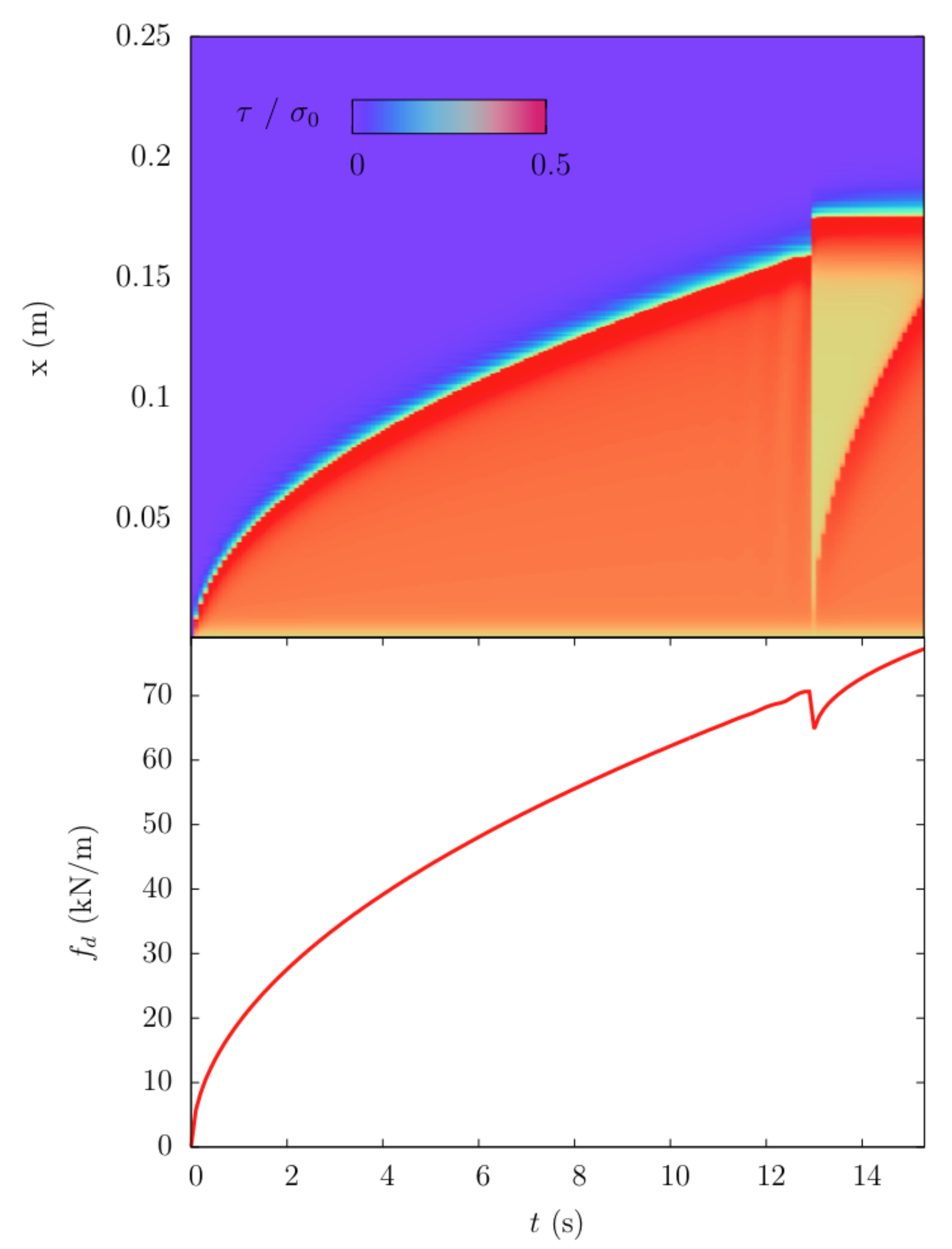}
  \caption{An example of a solution with the aging law, Eqs.~\eqref{eq:f}-\eqref{eq:phi}, for $H\!=\!0.005$ m. (upper panel) A space-time plot of the ratio between the interfacial shear stress and the applied normal stress $\tau/\sigma_0$. (lower panel) The corresponding overall applied force $f_d(t)$. Both panels clearly demonstrate the existence of a distinct instability occurring at $t\!\simeq\!13$ s. It corresponds to $L_c\!\simeq\!0.15$ m (cf.~Fig.~3 in the main text) and is followed by dynamic rupture propagation that occurs on a dramatically shorter timescale (hence it appears as a vertical step in the space-time plot in the upper panel).}\label{fig:spacetime}
\end{figure}

A homogeneous grid spacing with a typical number of $N_x\=400$ and $N_y\=150$ meshpoints in x- and y-direction, respectively, is used. An adaptive timestep is employed, which varies typically between $dt\=10^{-3}$ s and $dt\=10^{-6}$ s to capture both the slow penetration of the creep patch and faster rupture events that follow nucleation. The overall force (per unit depth) needed to maintain a fixed sideway velocity, $f_d(t)\=\!\int_0^H \sigma_{xx}(x\!=\!0, y, t)\,dy$, is tracked. As the instability that marks the onset of nucleation evolves much faster than the expanding creep patch, it is accompanied by an abrupt drop in $f_d(t)$, which provides a clear indication of an instability and hence is used to determine the critical nucleation length $L_c$ in a given simulation. An example, including the associated space-time plot of the ratio between the interfacial shear stress and the applied normal stress $\tau/\sigma_0$, is presented in Fig.~\ref{fig:spacetime}.

\begin{table}[h]
\centering
\begin{tabular}{ccc}
\hline
Parameter & Value & Units\\
\hline
$D$ & $5\times10^{-7}$ &m \\
  $b$ & $0.075$ & -\\
  $v^*$ & $10^{-7}$ & m/s\\
  $\theta$ & $5/18$ & -\\
  $\phi^*$ & $3.3\times10^{-4}$ & s\\
  $v_0$ & $10^{-9}$ & m/s\\
  $\xi$ & $0.005$ & -\\
\hline
\end{tabular}
\caption{The values of all of the friction law parameters appearing in Eqs.~\eqref{eq:f}-\eqref{eq:phi}.}
\label{tab:values}
\end{table}

\newpage

\end{document}